\documentclass[amssymb,aps,prli,twocolumn]{revtex4}
\usepackage{latexsym,amsmath,graphics,graphicx,cmbright,exscale,colordvi,accents,lipsum}
\usepackage[T1]{fontenc}
\usepackage{color}
\def\be{\begin{equation}}
\def\ee{\end{equation}}
\def\beq{\begin{eqnarray}}
\def\eeq{\end{eqnarray}}
\renewcommand{\Re}{\mathrm{Re}\,}
\renewcommand{\Im}{\mathrm{Im}\,}

\begin{document}
\renewcommand{\familydefault}{\sfdefault}
\renewcommand{\sfdefault}{cmbr}

\title{Transverse dynamical magnetic susceptibilities from regular {\it \bf static} density functional theory: 
Evaluation of damping and $g$-shifts of spin-excitations}
\author{Samir Lounis$^{\dagger}$}\email{s.lounis@fz-juelich.de}
\author{Manuel dos Santos Dias}
\author{Benedikt Schweflinghaus}
\affiliation{Peter Gr\"unberg Institut and Institute for Advanced Simulation, Forschungszentrum J\"ulich, 52425 J\"ulich \& JARA, Germany}

\begin{abstract}
The dynamical transverse magnetic Kohn-Sham susceptibility calculated within time-dependent density functional theory 
shows a fairly linear behavior for a finite energy window. 
This observation is used to propose a scheme where the computation of this quantity is greatly simplified. 
Regular simulations based on {\it static} density functional theory can be used to extract 
the dynamical behavior of the magnetic response function. Besides the ability to calculate elegantly 
damping of magnetic excitations, we derive along the way  useful equations giving the main 
characteristics of these excitations: effective $g$-factors and the resonance frequencies that can be 
accessed experimentally using inelastic scanning tunneling spectroscopy or spin-polarized electron energy loss spectroscopy.
\end{abstract}
\maketitle
\date{\today}

\section{Introduction}
Probing spin excitations on surfaces is a major focus of recently developed state of the art experimental techniques: 
spin polarized electron energy loss spectroscopy (SPEELS)~\cite{SPEELS1,SPEELS2} or inelastic scanning 
tunneling spectroscopy (ISTS)~\cite{heinrich,balashov,khajetoorians,pascual,otte,brune,brune2,hirjibehedin} 
allow, for example, to 
map partly the excited spectra of thin films and even nano-structures down to single adatoms on surfaces.
 All information on spin-excitations is encoded in the magnetic response of a material, 
but once probed experimentally it 
is dressed up differently depending on the measurement technique. 
 This explains the theoretical efforts, thriven by 
these experimental achievements, in developing methods that permit to simulate the dynamical transverse magnetic susceptibility, $\chi$, 
that, in linear response, describes the amplitude 
of the transverse spin motion produced by an external magnetic field of 
frequency $\omega$~\cite{costa,savrasov,katsnelson,buczek,sasioglu,lounis,costa_lounis_PRB,rousseau,lounis2}. 
  The excitations spectrum hinges on the 
details of the electronic structure of the probed material, 
 hybridization of electronic states and magnetization. 

Within SPEELS or ISTS, surface spin waves or localized spin excitations are excited via an exchange scattering process 
involving either a spin-polarized monochromatic electron beam or a tunneling current that interact with  
the sample. A gap in the dispersion curve of the spin waves (difficult to observe with actual SPEELS) or in the ISTS signal is 
the signature of symmetry breaking induced by spin-orbit coupling (SOC) or by an external magnetic field $B$. 
For single adatoms on non-magnetic substrates, the observed resonances, experimentally and 
theoretically, are located at the Larmor energy given by $g_{\mathrm{eff}}\mu_{\mathrm{B}}B$ with an effective spectroscopic splitting factor 
surprisingly different from 2~\cite{heinrich,balashov,khajetoorians,otte,brune,muniz,lounis,lounis2}. 
 The lifetime (damping) of the spin-excitations 
and $g$-shifts are the results of the coupling to Stoner (electron-hole pairs) excitations that change depending on the 
magnitude of the applied magnetic field~\cite{muniz,lounis,khajetoorians,schweflinghaus}. 
In fact, $g$-shifts were recognized since the early days of ferromagnetic 
resonance spectroscopy (FMR)~\cite{griffith} and were related to the presence of SOC~\cite{kittel,vanvleck,farle}. 
Contrary to FMR, ISTS probes the magnetic response locally 
instead of the full system. In this particular case, Mills and Lederer~\cite{mills1967} predicted that $g$ can be different 
from 2 even without SOC.  Indeed, during spin-precession, the surrounding bath of electrons are perturbed and spin-currents are emitted. 
The local response naturally differs from the global response as expected in the spin-pumping context.

To calculate dynamical magnetic quantities from ab initio, one uses either  
time-dependent density functional theory (TD DFT) ~\cite{TDDFT} or many-body perturbation theory (MBPT) coupled to DFT~\cite{MBPT}.  
Since the calculation of such quantities is a computational burden, their accessibility is limited~\cite{heil}. 
The goal of our manuscript is to present an attractive scheme wherein simple  equations derived within TD DFT allow 
to calculate dynamic properties just by computing {\it static} quantities 
defined at the Fermi energy; this is reasonable within linear response theory. Thus, our formulation can be implemented without tremendous 
efforts. 

A byproduct of our analysis is a simple formulation of 
the damping, local or non-local in space, of the spin-excitations. 
We note that the relation between the damping and the electronic structure has been predicted 
earlier~\cite{muniz,mills1967}. Moreover, several proposals (see e.g. Refs.~\cite{macdonald,damping1,hickey}) 
have been made to calculate the phenomenological damping parameter, $\lambda$, used 
to couple the precession of the moment to a reservoir when solving Landau-Lifshitz-Gilbert (LLG) equations~\cite{gilbert}. 
Calculations based on first-principles were also performed in 
Refs.~\cite{damping2,damping3,damping3-bis,damping4,damping5}, with a focus on $\lambda$ but not on the full magnetic response function. 

In our work, we derive  furthermore a  physically transparent form of the effective $g$-factor defining the response of the 
system to an external magnetic field. We relate its shift to the electronic structure and more precisely to the local density of states 
at the Fermi energy. It is interesting to note that  Qian and Vignale~\cite{vignale} recognized in their derivation of the spin-wave dynamics from TD DFT 
a dynamic term and a ``Berry curvature'' correction, which we believe can be connected to the $g$-shift mentioned earlier~\footnote{This topic is the subject of a future publication.}. 

We note that in the context of Kondo impurities $g$-shifts were also predicted many years ago~\cite{wolf}, which was recently used to intrepret experimental data extracted with 
ISTS~\cite{ternes,hirjibehedin,fernandez-rossier}. 
Using first-order perturbation theory in 
the weak coupling regime of a spin to the surrounding electronic bath, the shift in $g$ was shown to be proportional to $Jn(E_F)$, where J 
describes an $s$--$d$ type of interaction between a localized moment with the conduction electron and $n(E_F)$ is the density of states of conduction electrons 
at the Fermi energy. We stress that the latter is rather similar to our concept of renormalization of $g$ due to the electronic structure derived within TD DFT.   

\section{Method and framework}
\begin{figure*}
\begin{center}
\includegraphics*[angle=0,width=.8\linewidth]{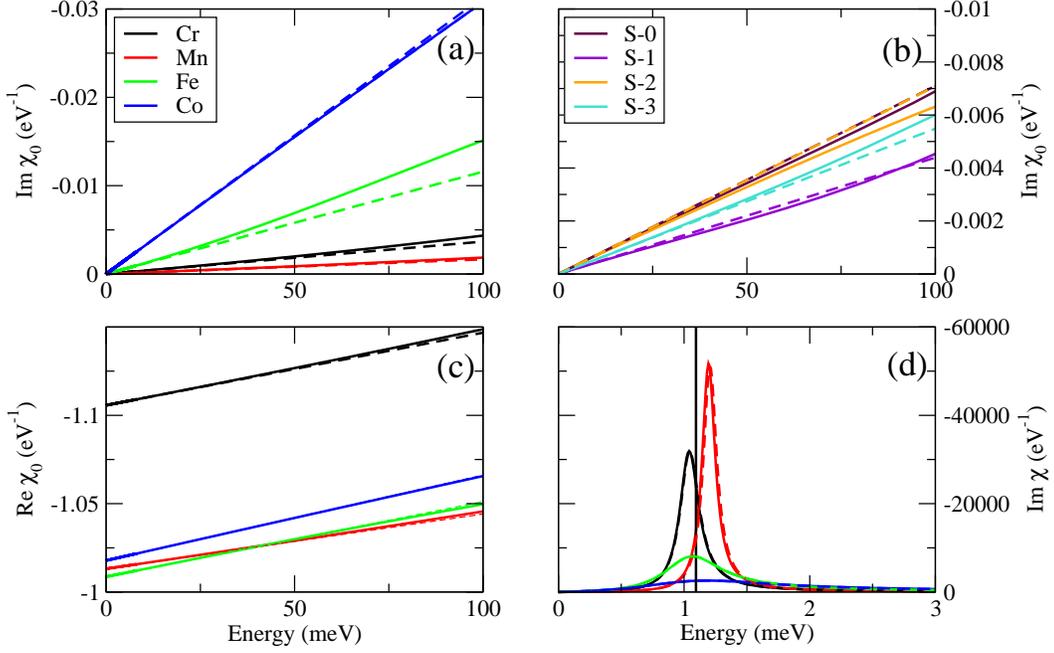}
\caption{The spherical component of the imaginary part of the transverse dynamic Kohn-Sham susceptibility, 
$\Im\chi_0$, shown for two systems: 4 different transition-metal adatoms on a Cu(111) surface (a), and 8 monolayers of Co on a Cu(001) surface, S-n denoting the n-th layer 
with respect to the surface (b). The 
results show a fair linear behavior up to large energies, which is well described by the dashed lines obtained with Eq.~\ref{important1}. As discussed in the text, a 
linear behavior is also expected for the real part of the Kohn-Sham susceptibility, which is indeed found as depicted in (c) for the case of transition-metal adatoms on 
a Cu(111) surface. After application of Eq.~\ref{important1} for the adatoms, the full transverse dynamic susceptibility is recovered (full lines versus dashed lines) in (d).}
\label{Chi0_fig}
\end{center}
\end{figure*}

The transverse dynamical magnetic susceptibility connecting the transverse magnetization to an infinitesimal transverse dynamical magnetic field 
is evaluated within TD DFT by solving the Dyson-like equation:
\beq
\chi = \frac{\chi_0}{1- U\chi_0} 
\label{Dyson}
\eeq
in a matrix notation. $U$ is the exchange and correlation kernel, which is assumed to be local in space and adiabatic.   
$\chi_0$ is the Kohn-Sham susceptibility whose imaginary part describes 
the density of Stoner excitations and can be written as a convolution of Green functions 
connecting the radial points $\vec{r}\,$ and $\vec{r}\,'$ centered respectively around sites $i$ and $j$:
\beq
\chi^{ij}_{0}(\vec{r}\,,\vec{r}\,';\omega)&=&-\frac{1}{\pi}\int dz f(z)
[G^{\downarrow}_{ij}(\vec{r}\,,\vec{r}\,';z+\omega)
\Im G^{\uparrow}_{ji}(\vec{r}\,',\vec{r}\,;z) \nonumber\\
&&
+ \Im G^{\downarrow}_{ij}(\vec{r}\,,\vec{r}\,';z)
G^{-\uparrow}_{ji}(\vec{r}\,',\vec{r}\,;z-\omega)]\label{chi0},
\eeq 
where $f(z)$ is the Fermi distribution function, $G$ and $G^-$ represent the retarded and advanced one-particle Green functions 
connecting atomic sites $i$ and $j$ and $\mathrm{Im}\,G=\frac{1}{2\mathrm{i}}(G-G^-)$. The assessment of $\chi_0$ requires 
to compute Green functions at energies $z$, $z + \omega $ and $z - \omega$. Also usual contour deformations when calculating the integral are not possible 
since there is a convolution of Green functions that are not analytical in the same complex half plane.

We evaluated the Kohn-Sham susceptibility for several systems utilizing our recently developed method~\cite{lounis,lounis2,schweflinghaus} based on the 
Korringa-Kohn-Rostoker Green function method~\cite{KKR}. In this scheme, the Green functions are projected on a chosen basis and the magnetic response to 
a spherically symmetric or site-dependent magnetic field is assumed. Within this approach $U$ 
simplifies to a single number per magnetic atom. 
In this case, the quantity of interest is the spherical part of the magnetic susceptibility 
$\overline{\chi}^{ij}({r},{r}';\omega)=\sum_{LL_1}\chi^{iLL_1;jL_1L}({r},{r}';\omega)$. 

In Figs.~\ref{Chi0_fig}(a),~\ref{Chi0_fig}(b) and~\ref{Chi0_fig}(c), we plot the imaginary and real parts of $\chi_0$ for Cr, Mn, Fe and Co single adatoms on Cu(111) and 
the imaginary part of $\chi_0$ for 8 Co monolayers on Cu(001). In the latter case, the layer-resolved quantity is shown for the four last Co surface layers. 
 Obviously, 
the imaginary and real parts of $\chi_0$ are rather linear with frequency.  
This linear regime defines the domain of applicability of the scheme proposed in this contribution, i.e.,  it is limited to few 
tens of meV up to a couple of hundred meV depending on the investigated material.
 
In this linear regime, the transverse  
Kohn-Sham susceptibility can thus be written in a Taylor expansion,
\beq
\chi_0 (\omega) = \Re\chi_0(0) + \left(\Re S + \mathrm{i} \Im S \right) \omega + \mathit{O}(\omega^2) \; ,
\label{linear_chi0}
\eeq 
where the slope $S$ is given by $\frac{\partial \chi_0}{\partial \omega}$ and 
$\Re\chi_0(0)$ is a Fermi-sea like term since it involves an energy integration up to 
the Fermi energy of a convolution of Green functions (see Eq.~\ref{chi0}). 
We demonstrate that the slopes defining the linearity of $\chi_0$ can be determined from one single 
DFT-based calculation. Once $\chi_0$ is known, 
the full-susceptibility $\chi$ can be evaluated after solving a Dyson-like equation (Eq.~\ref{Dyson}). 

For simplification, we assume that the systems of interest are magnetically collinear. 
Also, SOC is set aside allowing to decouple the longitudinal excitations from the transverse ones. 
However, one can use an external magnetic field to mimic the energy gap induced by SOC, 
which is of the order of the magnetic anisotropy energy. In other words, the magnetic field shifts 
the position of the resonances, thereby picking up electron-hole excitations that affect the corresponding lifetimes. 
In that case the gap is given by $g\mu_{\mathrm{B}}B$. Both procedures lead to 
about the same lifetimes~\cite{khajetoorians} within a tight-binding scheme~\cite{costa_lounis_PRB} or using TD DFT including SOC~\cite{dias}.

\section{Derivation and results}

 Since $G(z\pm\omega)=G(z)\mp G(z)\omega G(z\pm\omega)$ and taking the limit of $\omega \rightarrow 0$ we find that $ G(z\pm\omega) \simeq G(z)\mp\omega G(z)G(z)$ in matrix notation. 
We know that 
$\frac{\partial G(z\pm\omega)}{\partial \omega} \simeq \mp G(z) G(z) 
\simeq \pm \frac{d G(z)}{d z}$. Therefore for small $\omega$, 
\begin{widetext}
\beq
S^{ij}(\vec{r}\,,\vec{r}\,') 
 & =& \frac{\mathrm{i}}{2\pi }\Bigg[ \int^{E_F}{dz 
\Bigg(\frac{dG^{\downarrow}_{ij}(\vec{r}\,,\vec{r}\,';z)}{dz}G^{\uparrow}_{ji}(\vec{r}\,',\vec{r}\,;z)
+G^{*\downarrow}_{ji}(\vec{r}\,',\vec{r}\,;z)\frac{dG^{*\uparrow}_{ij}(\vec{r}\,,\vec{r}\,';z)}{dz}\Bigg)}
+G^{\downarrow}_{ij}(\vec{r}\,,\vec{r}\,';E_F)G^{*\uparrow}_{ij}(\vec{r}\,,\vec{r}\,';E_F)\Bigg].
\label{important1}
\eeq
\end{widetext}

Eq.~\ref{important1} is among the important results reported in this paper. It shows that extracting the linear behavior of the susceptibility requires an integral along a complex energy contour of 
a convolution of Green functions that are analytical on the same complex half plane. Therefore usual techniques used in regular methods based on Green functions can be applied (see e.g. Ref.~\cite{KKR}). Also the 
evaluation of $dG/dz$ is numerically straightforward and the 
integral over the energy is 
evaluated only once, thus the computational costs 
are dramatically reduced compared to those needed for the evaluation of Eq.~\ref{chi0}. 

We utilized Eq.~\ref{important1} to evaluate the Kohn-Sham susceptibilities for the cases mentioned earlier, 
for instance Cr, Mn, Fe and Co adatoms on Cu(111) surface and Co thin film on Cu(001) surface. The results are 
shown in Fig.~{\ref{Chi0_fig}} as dashed lines, which fit rather well the overall behavior of $\chi_0$. 
For indication, the slopes obtained for the adatoms are listed in Table~\ref{table}. In 
Fig.~{\ref{Chi0_fig}}(d), we plot the imaginary part of the full transverse dynamical susceptibility using $\chi_0$ obtained with the regular scheme, i.e. integral 
given by Eq.~\ref{chi0} (full lines) and with Eqs.~\ref{linear_chi0} and~\ref{important1} (dashed lines).
Also  values of $g_{\mathrm{eff}}$ obtained with both type of calculations are presented in Table~\ref{table}.  The agreement is extremely good which validates our proposal. 

For the adatom case and in the linear regime,  
 the imaginary part 
of $\chi$ given by 
\beq 
\Im \chi = \frac{\Im\chi_0}{[1-U\Re\chi_0]^2+[U\Im\chi_0]^2}
\eeq
 can be rewritten as   
\beq
\Im\chi = \frac{\omega \Im S}{\left[1-U(\omega\Re S  +\Re\chi_0(0))\right]^2 + (U\omega \Im S)^2} \; ,
\label{dyson3}
\eeq
which defines the density of magnetic excitations, i.e. it gives the theoretical spectrum corresponding 
to the experimental measurements. 

In Eq.~\ref{dyson3} one recognizes similarities with the regular Dyson equations 
allowing to compute Green functions. For example, an orbital originally located at $E_0$, broadens and experiences a shift 
after hybridization with the electronic background. The latter can be described via 
the self-energy $\Sigma$ and the corresponding imaginary part of the Green function reads 
\beq
\Im G = \frac{\Im \Sigma}{\left([E-E_0-\Re\Sigma]^2+[\Im\Sigma]^2\right)}.
\label{Sigma}
\eeq
One deduces that $\chi_0$ acts as a 
self-energy with an imaginary part defining the lifetime of the excitations described by $\chi$ 
while the real part of $\chi_0$ defines its energy-shift. 

\begin{table}
\begin{tabular}{l| c c c c}
\hline
\hline
       & Cr & Mn & Fe & Co \\
\hline
$\Re S$ (eV$^{-2}$) & -0.411 & -0.310 & -0.425 & -0.483 \\   
$\Im S$ (eV$^{-2}$) & -0.037 & -0.017 & -0.116 & -0.313 \\
$\beta$ (eV$^{-2}$) & 0.784 &  0.681 &  0.858 & 1.237 \\
$g_{\mathrm{eff}}$ & 1.89 & 2.17 & 1.95 & 2.14 \\
$g_{\mathrm{eff}}$ (after using Eq.~\ref{important1}) & 1.90 & 2.19 & 1.95 & 2.15 \\
\hline
\hline
\end{tabular}    
\caption{\label{table} Values of the slope of the Kohn-Sham susceptibility, $\Re S$ and $\Im S$, 
at small frequencies calculated with Eq.~\ref{important1} for different adatoms on 
Cu(111) surface. Those values lead to a good description of the low-energy regime of the Kohn-Sham susceptibility and of the full transverse susceptibility as can be observed in Fig.~\ref{Chi0_fig}. 
 For instance, values of $g_{\mathrm{eff}}$ obtained from both schemes are about the same.}
\end{table}

\subsection{Imaginary part of the slope of $\chi_0$}
It is interesting to note that the imaginary part of the slope $S$ can be deduced in a transparent form that reads
\beq
\Im S^{ij}(\vec{r}\,,\vec{r}\,')&=&
-\pi n_{ij}^{\downarrow}(\vec{r}\,,\vec{r}\,';E_F) n_{ji}^{\uparrow}(\vec{r}\,',\vec{r}\,,E_F) \; , \quad
\label{imag_chi0}
\eeq
where $n_{ij}^\sigma(\vec{r}\,,\vec{r}\,';E)=-\frac{1}{2\pi i}(G_{ij}^\sigma(\vec{r}\,,\vec{r}\,';E)-G_{ij}^{*\sigma}(\vec{r}\,,\vec{r}\,';E))$ 
and $\sigma\in\{\uparrow,\downarrow\}$.   Eq.~\ref{imag_chi0} is compelling and useful: 
the imaginary part of the change of $\chi_0$ at small frequencies is related to the density matrix for 
each spin channel at the Fermi energy. $\Im S$ can thus be considered as a Fermi-surface-like term. 
If $\vec{r}\,=\vec{r}\,'$, i.e., the susceptibility is local in space  and depends on the product of 
energy-dependent charge density of opposite spin-character. 
This is the second important result of this work. The lifetime of the excitation and, thus, 
the damping can be calculated exactly after a one-shot calculation from a regular static DFT run 
without even having to perform an energy integration. 
We propose to use this damping directly on top of the adiabatic 
approach usually utilized to extract the magnetic exchange interactions (see e.g. Refs.~\cite{lichtenstein,lounis3,szilva}) and the 
related dispersion of spin-waves or the localized spin-excitations spectra (e.g. Refs.~\cite{bergqvist,bauer,boettcher}). However, 
one has to keep in mind that Stoner excitation will impact on the position of spin waves resonances via the real part of 
the Kohn-Sham susceptibility, which will be the topic of the next paragraph. 

We note that Eq.~\ref{imag_chi0} reduces to 
\beq
\Im \overline{S}^{ij} &=&
-\pi \sum_{LL_1} n^{ij\downarrow}_{LL_1}(E_F) n^{ji\uparrow}_{L_1L}(E_F)\label{imag_chi0_simple}
\eeq 
if we use the framework presented in Ref.~\cite{lounis}, where a 
projection of the Green functions on the $d$ states is made and the response to 
a spherically symmetric or site-dependent magnetic field is assumed.

Eqs.~\ref{imag_chi0} or~\ref{imag_chi0_simple} have paramount physical implications: 
The way the minority or majority spin-levels intersect the Fermi energy defines the damping of 
the excitations in terms of the product of density of states of opposite spin character. 
This explains for example that Fe and Co adatoms on several surfaces (e.g. Cu(001), Cu(111) and Ag(111) surfaces) have broad resonances in 
their excitation spectra since their minority-spin local density of states experiences a resonance 
close to the Fermi energy contrary to Mn and Cr adatoms on the same surfaces, which have much sharper resonances 
because the Fermi level lies between the majority and minority states~\cite{lounis,schweflinghaus,khajetoorians}. 
Considering that scanning tunneling microscopy can be used to map the spin-polarized local density 
of states at the Fermi energy, an experimental estimate of the damping can be provided via Eq.~\ref{imag_chi0}. 

\subsection{Real part of the slope of $\chi_0$} 
Compared to the imaginary part, the real part of $S$ is, however, not easy to simplify. To 
obtain a simple form of the real part, both sides of Eq.~\ref{important1} are multiplied 
by the exchange splitting potential, $B^i$ from the right and $B^j$ from the left, integrated 
over $d\vec{r}\,$, $d\vec{r}\,'$, and summed up over   $i$ and $j$, where $B^i$ is given by 
the difference between the potentials of each spin channel $(V^{\uparrow}-V^{\downarrow})$ for atom $i$. 
We make use of $G^{\uparrow}=G^{\downarrow}+G^{\downarrow}B{G^{\uparrow}}$ and define the total moment of the system 
$M$ in order to reduce 
$\sum_{ij}\int d\vec{r} d\vec{r}\,' B^i(\vec{r}\,)S^{ij}(\vec{r}\,,\vec{r}\,') B^j(\vec{r}\,')$ (at omega=0) 
to a sum rule:
\beq
\sum_{ij}B^i S^{ij} B^j &=&
-M + \sum_jG^{*\uparrow}_{jj}(E_F)B^j-\sum_i B^iG^{\downarrow}_{ii}(E_F)
\nonumber\\
&+&\frac{1}{2\pi \mathrm{i}} B^i
(G^{\downarrow}_{ij}(E_F)G^{*\uparrow}_{ij}(E_F))B^j,\label{important2}
\eeq
in a matrix notation. Interestingly, the calculation of the total magnetic moment 
involves an energy integration up to the Fermi energy and thus 
can be identified as a Fermi-sea-like contribution to the sum rule while the remaining terms are  
Fermi-surface-like contributions.


In the case of a single magnetic adatom deposited on a non-magnetic substrate 
and within the spherical approximation defined earlier, Eq.~\ref{important2} becomes
\beq
 \overline{S}^{ii} &=& -\frac{1}{B_i}M_i\frac{1}{B_i}
-\frac{1}{2\pi i}
\sum_L[\frac{1}{B^i}G^{*\uparrow}_{ii,LL}(E_F)-G^{\downarrow}_{ii,LL}(E_F)\frac{1}{B^i}]
\nonumber\\
&&+\frac{1}{2\pi i} 
\sum_{LL_1}(G^{\downarrow}_{ii,LL_1}(E_F)G^{*\uparrow}_{ii,L_1L}(E_F)),
\eeq
and for instance, the real part is given by
\beq
 \Re \overline{S}^{ii} &=& -\frac{1}{B^i}M^i\frac{1}{B^i}
-\frac{n^{i}(E_F)}{2B^i}\nonumber\\
&+&\Im\Big[ \frac{1}{2\pi} 
\sum_{LL_1}(G^{\downarrow}_{ii,LL_1}(E_F)G^{*\uparrow}_{ii,L_1L}(E_F)) \Big],\label{real_chi0_simple}
\eeq
where we have replaced $\Im \frac{1}{\pi }
\sum_L[G^{*\uparrow}_{ii,LL}(E_F)-G^{\downarrow}_{ii,LL}(E_F)]$ by the total density of states at the Fermi energy 
$n^{i}(E_F)$. $ \Re S$  depends on properties defined at the Fermi energy, Fermi-surface-like terms, 
and on an energy integrated  
term, the magnetic moment, which defines a Fermi-sea-like contribution. Unlike  $\Im S$, $\Re S$ is in this particular case a 
complicated combination of parameters. It hinges on the magnetization, the exchange splitting, the density of states at the Fermi energy as 
well as the real part of a product of Green functions of opposite spin.

\subsection{Resonance energy, $g$--values and spin-excitation's lifetime of single adatoms}
In this model, the resonance energy for the single adatom case, $\omega_{\mathrm{res}}$, in the 
excitation spectrum is given by the extremum of $\Im\chi$, {i.e.} $\frac{\partial\Im\chi}{\partial \omega}= 0$. From Eq.~\ref{dyson3} we find 
\beq
\omega_{\mathrm{res}} = \frac{\vert\frac{1}{U}- \Re\overline{\chi}_0(0)\vert}{\sqrt{\Re S^2+\Im S^2}} \; .
\eeq
Thus, the position of the resonance depends equally on the slope $S$  as well as on the value of 
$\Re\overline{\chi}_0$ at $\omega = 0$. 
Furthermore, the full-width half maximum value, which defines the spin-excitation's lifetime, turns out to be proportional to $\omega_\mathrm{res}$ 
and reads
\beq
\mathrm{FWHM} = 2 \cdot \sqrt{\left(2+\frac{\Re S}{\sqrt{(\Re S)^2+(\Im S)^2}}\right)^2-1} \cdot \omega_\mathrm{res} \; .
\eeq
If no external magnetic field is applied along the $z$ direction, $\Re\overline{\chi}_0(0)=1/U$ since the 
Goldstone mode has to be satisfied and consequently $\omega_{\mathrm{res}} =0$.
If a magnetic field $B$ is applied along the $z$ direction, the effective Larmor energy is given by 
$\omega_{\mathrm{res}} = g_{\mathrm{eff}}\mu_BB $ 
and
\beq
g_{\mathrm{eff}}&=& \frac{1}{\mu_BB} \frac{\vert\frac{1}{U}- \Re\overline{\chi}_0(0)\vert}{\sqrt{(\Re S)^2+(\Im S)^2}}
\eeq
describes the response of the system to this field. Results obtained from the previous equation recover naturally the values extracted from 
Fig.~\ref{Chi0_fig}(d) and shown in Table~\ref{table}.

Using once more a Tailor expansion of $\chi_0$ in terms of the applied magnetic field and the 
sum rule relating $\chi_0$ to $U$ at zero-magnetic field we end up with:
\beq
g_{\mathrm{eff}}&\approx& g \sqrt{\frac{\beta^2}{(\Re S)^2+(\Im S)^2}} \label{g-shift}
\eeq
where $\beta$ is given by the first two terms of the right-hand side of Eq.~\ref{real_chi0_simple}, i.e., it can be 
written as 
\beq
\beta = -\frac{1}{B^i}M^i\frac{1}{B^i}-\frac{n^{i}(E_F)}{2B^i}.
\eeq
 This result has important consequences. If the local density of states at the Fermi energy vanishes, 
$\Im S$ goes to zero while $\beta$ converges to $\Re S$ leading to $g_{\mathrm{eff}}=2$. However, 
any finite occupation at the Fermi energy, in other words, any metallic behavior will shift $g$ away from 2. Access to the exchange 
splitting and the magnetic moment of a material 
may help to estimate the shift in $g$ as well as the lifetime of spin-excitations via Eq.~\ref{g-shift}. Values of $\beta$ for Cr, 
Mn, Fe and Co adatoms are presented in Table~\ref{table}.

Besides $g$, another quantity describes the spin-dynamics, namely the damping parameter $\lambda$ encountered in the 
LLG equations \cite{gilbert}. Several studies were recently devoted to the formulation and computation of the Gilbert damping parameter from 
realistic electronic structures~\cite{damping1,damping2,damping3,damping3-bis,damping4,damping5}. Here we use Eq.~\ref{imag_chi0} and 
as an example, we derive the form of $\lambda$ considering the particular case of 
a single magnetic impurity. After applying a transverse time-dependent magnetic field $\vec{B}_t$ on top of the static field $\vec{B}$ parallel to $\vec{M}$, we linearize the LLG equations:
\beq 
\frac{d\vec{M}}{dt}=-\frac{g\mu_B}{\hbar} \vec{M}\times(\vec{B}+\vec{B}_t)-\lambda\frac{g\mu_B}{\hbar|M|}\vec{M}\times(\vec{M}\times(\vec{B}+\vec{B}_t))
\eeq
 and relate the induced 
transverse magnetization to the transverse field via a dynamical transverse susceptibility in a fashion similar to TD DFT:
\beq
\chi&=&-\frac{g{\mu_B}(1+\mathrm{i}\lambda)M}{{\omega} - g{\mu_B}B -\mathrm{i}\lambda g{\mu_B}B}
\eeq
whose imaginary part can be compared to Eq.~\ref{dyson3} in order to identify $\lambda$. We find that 
\beq
\lambda = \Im S,
\eeq 
which leads to a formulation of the damping parameter entering the LLG equations. From the previous equation, one deduces once more that a metallic system leads to a finite value 
of $\lambda$.

\section{Conclusion}
In this paper we have derived a simple formula that allows to compute the slope of the dynamical 
transverse magnetic susceptibility from static DFT calculations. The cost of calculations are then tremendously reduced in comparison 
with the usual method that permits the calculation of the magnetic response function. We provided a comparison between the two methods and demonstrate their  mutual 
agreement for a reasonable energy window. We applied our formalism to a Co thin film on Cu(111) surface and 
to 3$d$ adatoms on the same substrate.   
The Fermi surface defines the dynamical behavior, which is a reasonable statement within linear response theory. 
Furthermore, our formulation permits a straightforward interpretation of the origin of damping observed in the measured 
spin-excitations spectra and the dependence on the electronic structure. 
Using a simple model, we derived an analytical form for the effective $g$ value of a single magnetic adatom on a non-magnetic substrate and we explain thereby the origin of 
its shift by the degree of metallicity or itinerancy. Finally, we suggested a mapping procedure from ab initio to a Heisenberg model used when solving LLG equations. 
For instance, a form of the Gilbert damping parameter due to Stoner excitations is introduced, which is in agreement with previous proposals.

\section*{Acknowledgments}
We acknowledge discussions with late D.~L.~Mills, A.~T.~Costa, P.~H.~Dederichs and S.~Bl\"ugel. Research supported by 
the HGF-YIG Programme VH-NG-717 (Functional nanoscale structure and probe simulation laboratory -- Funsilab).

\end{document}